\documentclass[twocolumn,english,reprint, longbibliography, superscriptaddress, breaklinks=true, showkeys, showpacs=false, nofootinbib]{revtex4}
\usepackage[T1]{fontenc}
\usepackage[latin9]{inputenc}
\usepackage{color}
\usepackage{babel}
\usepackage{amsmath}
\usepackage{amssymb}
\usepackage{subfigure}
\usepackage{graphicx}
\usepackage{physics} 
\usepackage{hyperref}
\hypersetup{
    colorlinks=true,
    linkcolor=blue,
    filecolor=blue,      
    urlcolor=blue,
    }
\makeatletter
\@ifundefined{textcolor}{}
{%
 \definecolor{BLACK}{gray}{0}
 \definecolor{WHITE}{gray}{1}
 \definecolor{RED}{rgb}{1,0,0}
 \definecolor{GREEN}{rgb}{0,1,0}
 \definecolor{BLUE}{rgb}{0,0,1}
 \definecolor{CYAN}{cmyk}{1,0,0,0}
 \definecolor{MAGENTA}{cmyk}{0,1,0,0}
 \definecolor{YELLOW}{cmyk}{0,0,1,0}
}
\pdfoutput=1
\hypersetup{colorlinks=true,citecolor=blue,linkcolor=cyan,urlcolor=blue,filecolor= green, breaklinks=true}
\usepackage{url}
\usepackage{breakurl}
\makeatother
\begin{document}

\title{Simulating noisy quantum channels via quantum state preparation algorithms}

\author{Marcelo S. Zanetti}
\address{Electronics and Computing Department, Technology Center, Federal University of Santa Maria, Roraima Avenue 1000, Santa Maria, Rio Grande do Sul, 97105-900, Brazil}

\author{Douglas F. Pinto}
\address{Physics Department, Center for Natural and Exact Sciences, Federal University of Santa Maria, Roraima Avenue 1000, Santa Maria, Rio Grande do Sul, 97105-900, Brazil}

\author{Marcos L. W. Basso}
\address{Center for Natural and Human Sciences, Federal University of the ABC, States Avenue 5001, Santo Andr\'e, S\~ao Paulo, 09210-580, Brazil}

\author{Jonas Maziero}
\email[Corresponding author: ]{ jonas.maziero@ufsm.br}
\address{Physics Department, Center for Natural and Exact Sciences, Federal University of Santa Maria, Roraima Avenue 1000, Santa Maria, Rio Grande do Sul, 97105-900, Brazil}

\begin{abstract}
In Refs. [Phys. Rev. A 96, 062303 (2017)] and [Sci. China Phys. Mech. Astron. 61, 70311
(2018)], the authors reported an algorithm to simulate, in a circuit-based quantum computer, a general quantum channel (QC). However, the application of their algorithm is limited because it entails the solution of intricate non-linear systems of equations in order to obtain the quantum circuit to be implemented for the simulation. Motivated by this issue, in this article we identify and discuss a simple way to implement the simulation of QCs on any $d$-level quantum system through quantum state preparation algorithms, that have received much attention in the quantum information science literature lately. We exemplify the versatility of our protocol applying it to most well known qubit QCs, to some qudit QCs, and to simulate the effect of Lorentz transformations on spin states. We also regard the application of our protocol for initial mixed states. Most of the given application examples are demonstrated using IBM's quantum computers.
\end{abstract}


\maketitle
 

\section{Introduction} 

One of the fundamental postulates of quantum mechanics asserts that the evolution of the state of a closed quantum system $A$ is described by a unitary operator $U$ acting on a Hilbert space $\mathcal{H}_A$, i.e., $U^\dagger U=UU^\dagger=\mathbb{I}$ with $\mathbb{I}$ being the identity operator in $\mathcal{H}_A$, while $U$ is derived from the system's Hamiltonian $H_t$ through the solution of the Schr\"odinger equation $i\hbar\partial_{t}U=H_t U$ at time $t$. 
If the closed system is initially prepared in the state $\rho_A$, it evolves to the state $\rho_A'=U\rho_AU^\dagger$. 
Further research showed that the evolution of open quantum systems can be more generally represented through quantum operations, that are defined by a set of Kraus operators $\{K_j\}$ in $\mathcal{H}_A$ satisfying the completeness relation $\sum_jK_j^\dagger{}K_j=\mathbb{I}$.
Considering this general setting, if the system is prepared in the state $\rho_A$, it will evolve to the state $\rho_A'=\sum_jK_j\rho_AK_j^\dagger{}$, also known as the Kraus' operator-sum representation for quantum operations  \cite{kraus}.
One of the important implications of the completeness restriction is that it results in this map being completely positive and trace preserving (CPTP), ensuring that the evolved state is represented by a valid density matrix. 
A further generalization can be adopted by permitting that $\sum_jK_j^\dagger{}K_j\leq\mathbb{I}$, where the inequality does not affect the previously established complete positivity, but will result in a non-trace preserving map, allowing, for instance, for the description of selective quantum measurements \cite{Nielsen,Wilde}.


In Quantum Information Science (QIS), quantum operations that are CPTP are referred to as Quantum Channels (QCs) \cite{christandl09,Weedbrook11,iten17}. 
Any physical process that maps an initial state to a final state can be modelled as a QC, with examples being found in many application areas, such as: quantum thermodynamics \cite{binder15,barra17,oliveira20},  quantum gravity \cite{matsumura22}, quantum theory \cite{torre12}, quantum information and communication \cite{christandl09,berta13,caruso14,bromley15,paz-silva19}, and quantum computation and algorithms \cite{aharonov93,farhi98,kemp20,dong21,lewenstein21,costa21}.
In general, with the QCs framework, state evolution can be viewed as state perturbation and this is a suitable description of a system evolving through its interaction with a noisy environment.
Therefore, QCs can model how quantum systems behave in the presence of different noise and error sources \cite{Nielsen,oliveira20}.


The progress in important areas, such as the highly anticipated quantum computers and quantum simulators, strongly depends on a precise description of noisy QCs. 
In the case of quantum computers, the most successful implementations rely on a cryogenic chamber mitigating the coupling with the environment to emulate a closed system \cite{ibmq}.
Therefore, in order to characterize the limitations of an implementation, for example, to estimate decoherence times \cite{zurek03}, or to design error correcting codes \cite{Nielsen}, a quantum computer need to be considered as an open quantum system, with its state evolving according to noisy QCs models, and this can be achieved by simulating the  effects of QCs with quantum experiments.

It is worth to mention that there are several ways one can describe open systems dynamics, as for example the time discrete Kraus' operator sum representation \cite{kraus}, the time continuous Lindblad master equation \cite{lindblad}, the stochastic Schr\"odinger equation \cite{sschrodinger}, etc. There are also many works in the literature that deal with the simulation of open system dynamics, with a variety of techniques applied. But as it is not our objective to make a complete review of the subject here, we refer the interested reader to Refs. \cite{carlo,wang,fisher,wang2,cialdi,lu,han,schlimgen,kamakari,schlimgen2,schlimgen3,salles,david,xin,wei}. 

For the Kraus' representation we use in this article, the simulation approach goes as follows. An open quantum system $A$ and its environment $B$ are viewed as a closed quantum system subjected to unitary evolution according to the dynamics dictated by a given QC model.
By taking the partial trace over $B$, i.e. by discarding $B$, this evolution can be represented by an isometric map $V_{AB}$.
Formally, a noisy QC with Kraus' operators $\{K_{j}\}$ can be implemented coherently through an isometric transformation \cite{Wilde}
\begin{equation}
V_{AB}|\psi\rangle_{A}\otimes|0\rangle_{B} = \sum_{j}(K_{j}|\psi\rangle_{A})\otimes|j\rangle_{B},
\label{eq:iso}
\end{equation}
with $|\psi\rangle_A$ being the initial state of the system $A$ and $\{|j\rangle_B\}$ is an orthonormal basis for the system $B$.

Simulations of quantum channels generally start from Eq.~(\ref{eq:iso}). In Refs.~\cite{wei,xin}, the authors followed this path and reported the following algorithm to simulate a QC. The initial state of the main and auxiliary systems is set to $|\Psi_0\rangle_{AB}=|\psi\rangle_A \otimes|0\rangle_B$. Then, a unitary operation $V$ is applied to system $B$. After that, one implements the following controlled unitary operation: $U_c = \sum_{j=0}^{d-1}U_j \otimes|j\rangle_B \langle j|$, where $U_j$ are unitary operations acting in $\mathcal{H}_A$ and $d=d_A^2$ with $d_A =\dim\mathcal{H}_A$. Finally, the unitary operation $W$ is applied on the subsystem $B$. With this algorithm, the state on the right hand side of Eq.~(\ref{eq:iso}) is prepared with
\begin{equation}
    K_j = \sum_{l=0}^{d-1}W_{j,l}V_{l,0}U_l.
\end{equation}
If $U_j$ is an operator basis for operators acting on $\mathcal{H}_A$, then this protocol is seen to implement any QC on the system $A$. However, given a QC, i.e., given $\{K_j\}$, besides having to transform the unitaries $\{U_j\}$ into elementary gates of the quantum computer, one has to solve an intricate nonlinear system of equations in order to determine $V,W,\{U_j\}$, that, by their turn, determine the quantum circuit to be utilized for the quantum simulation.

Algorithms for quantum state preparation (QSP) have been used as subroutines for accomplishing many tasks \cite{kitaev,shende,plesch,arrazola,araujo,he,zhang,veras}, as for instance for implementing the general quantum Fourier transform \cite{kitaev}. Motivated by the issues just discussed about the quantum channel simulation algorithm of Refs.~\cite{wei,xin}, in this article we report a simple protocol that implements quantum channels on any discrete quantum system $A$, with a Hilbert space $\mathcal{H}_{A}$, via QSP algorithms. 
For any $d_A$ and any number of elements of the QC, if the Kraus' operators $K_j$ are known, in principle it is possible to calculate the right hand side of Eq.~(\ref{eq:iso}): \begin{equation}
|\Psi\rangle_{AB} = \sum_{j}\big(K_{j}|\psi\rangle_{A}\big)\otimes|j\rangle_{B}.
\label{eq:psiAB}
\end{equation}
Once obtained this vector, we can use QSP algorithms to prepare it. Afterwards, we ignore the auxiliary system $B$ and perform quantum state tomography of the system $A$ state. 
It is worthwhile to mention that if one is interested in simulating the individual action of each Kraus' operator, this can also be done by measuring the ancilla system $B$ in the basis $\{|j\rangle\}$ and post-selecting the results.

The remainder of this article is organized as follows. In Sec.~\ref{sec:qubit}, we describe our protocol in details by analysing a few interesting and recurrent two-level, i.e. one-qubit, noisy QCs. In Sec.~\ref{sec:qudit}, we apply our protocol to one-qudit QCs. In Sec.~\ref{sec:lorentz}, we show that our approach can be employed to simulate an interesting kind of QC, the one generated as consequence of the Lorentz transformations. Finally, in Sec.~\ref{sec:mixed}, we generalize our protocol to handle mixed initial states. In Sec.~\ref{sec:conc}, we present our final remarks.


\section{One-qubit noise channels}
\label{sec:qubit}

In this section, we start presenting a novel protocol for QCs simulation based on quantum state preparation algorithms. 
The protocol is implemented through the following steps:
\begin{enumerate}
    \item Define the QC model to simulate a noisy environment via a set $\{K_j\}$ of Kraus' operators;
    \item Prepare a quantum system $A$ with initial state $|\psi\rangle_A$; 
    \item Prepare an ancilla system $B$ on state $|0\rangle_B$, with $B$ having $n=|\{K_j\}|$ levels, or $\mathcal{O}(\log_2 n)$ qubits;
    \item Compute the state vector of Eq. (\ref{eq:psiAB}).
\item Prepare $|\Psi\rangle_{AB}$ using quantum state preparation techniques;
\item Discard $B$ and analyse the reduced dynamics of $A$, e.g. with quantum state tomography (QST). 
\end{enumerate}

It is worthwhile mentioning that, in principle, the QST part of our method is similar to that of other QC simulation methods. Here we applied the ready to use function of Qiskit \cite{qiskit}, that is based on the maximum likelihood QST method \cite{banaszek,james}. However, there is a large body of literature on QST and several other methods were produced and could be utilized as well (see e.g. \cite{gross,faist,gupta,koutny,aaronson,nguyen}).   

In the following, we illustrate the application of our protocol to simulate noisy QCs applied to one-qubit quantum systems, and we make a few quantum simulations on circuit-based quantum computers available at IBMQ \cite{ibmq}. 
For testing our protocol, we shall consider the dynamics of quantum coherence, a distinctive property of quantum mechanics and an important quantum information processing resource \cite{Nielsen,peng16}, under the action of different quantum channels. 
In our investigations, we adopt the $l_1$-norm of coherence \cite{baumgratz2014,hu18}:
\begin{equation}
C_{l_1}(\rho) = \sum_{j\ne k}|\rho_{j,k}|,
\end{equation}
where $\rho_{j,k}$ are the matrix elements of the density operator $\rho$ when represented in a certain reference basis, that here we consider as being the computational basis. For one qubit systems, this basis is denoted by $\{|0\rangle,|1\rangle\}$. Besides, it is worth mentioning that $\rho$ is the reduced density operator of the quantum system $A$ after discarding the ancilla $B$ in Eq.~\eqref{eq:psiAB}.



Let us discuss the noisy QC models that will be used to apply our protocol. We start with the Pauli's quantum channel (PQC) \cite{pauliacm}, which generalizes multiple QC models well known in the literature.
Mathematically, it is defined as
\begin{equation}
\Lambda_{p}(\rho_A)=\sum_{j,k=0}^{1}p_{j,k}Z^{j}X^{k}\rho_A  X^{k}Z^{j},
\label{eq:pauli}
\end{equation}
where $X$ and $Z$ are Pauli's matrices,
$p_{j,k}$ is a probability distribution, and $\sqrt{p_{j,k}}Z^{j}X^{k}=K_{j,k}$ are the Kraus' operators for the PQC. 
One can easily recast the PQC as follows
\begin{align}
& \Lambda_{p}(\rho_A) = \Lambda_{p}^{p_I,p_X,p_Z,p_Y}(\rho_A) \nonumber \\
& = p_I \rho_A + p_X X\rho_A X + p_Z Z\rho_A Z + p_Y Y\rho_A Y.
\end{align}
With this, one can represent the following noisy QCs as particular cases:
\begin{itemize}
\item Bit flip: $\Lambda_{bf}\equiv\Lambda_{p}^{1-p,p,0,0}$,
\item Phase flip: $\Lambda_{pf}\equiv\Lambda_{p}^{1-p,0,p,0}$,
\item Bit-Phase flip: $\Lambda_{bpf}\equiv\Lambda_{p}^{1-p,0,0,p}$,
\item Depolarizing: $\Lambda_{d}\equiv\Lambda_{p}^{(4-3p)/4,p/4,p/4,p/4}$.
\end{itemize}

For a general pure initial one-qubit state $|\psi\rangle_A = \cos(\theta/2)|0\rangle + e^{i\phi}\sin(\theta/2)|1\rangle$, with $\theta\in[0,\pi]$ and $\phi\in[0,2\pi)$, one can use our protocol to simulate any PQC by preparing the state
\begin{align}
& |\Psi\rangle_{AB} = \sqrt{p_I}\mathbb{I}|\psi\rangle_{A}\otimes|0\rangle_{B} + \sqrt{p_X}X|\psi\rangle_{A}\otimes|1\rangle_{B} \\
& + \sqrt{p_Z}Z|\psi\rangle_{A}\otimes|2\rangle_{B} + \sqrt{p_Y}Y|\psi\rangle_{A}\otimes|3\rangle_{B} \nonumber \\
& = \sqrt{p_{I}}\cos(\theta/2)|000\rangle_{Abc} + \sqrt{p_{I}}e^{i\phi}\sin(\theta/2)|100\rangle_{Abc} \nonumber \\
& + \sqrt{p_{X}}\cos(\theta/2)|101\rangle_{Abc} + \sqrt{p_{X}}e^{i\phi}\sin(\theta/2)|001\rangle_{Abc} \nonumber \\
& + \sqrt{p_{Z}}\cos(\theta/2)|010\rangle_{Abc} - \sqrt{p_{Z}}e^{i\phi}\sin(\theta/2)|110\rangle_{Abc} \nonumber \\
& + i\sqrt{p_{Y}}\cos(\theta/2)|111\rangle_{Abc} - i\sqrt{p_{Y}}e^{i\phi}\sin(\theta/2)|011\rangle_{Abc}, \nonumber
\end{align}
where the qubits $b$ and $c$ represent the auxiliary system $B$.
In remainder of this section, we consider the qubit prepared initially in the following state of maximal coherence: 
\begin{equation}
|\psi\rangle_A=\frac{1}{\sqrt{2}}(|0\rangle+|1\rangle) \equiv |+\rangle_A.
\end{equation}

\begin{figure}[t!]
\centering
\includegraphics[width=0.53\textwidth]{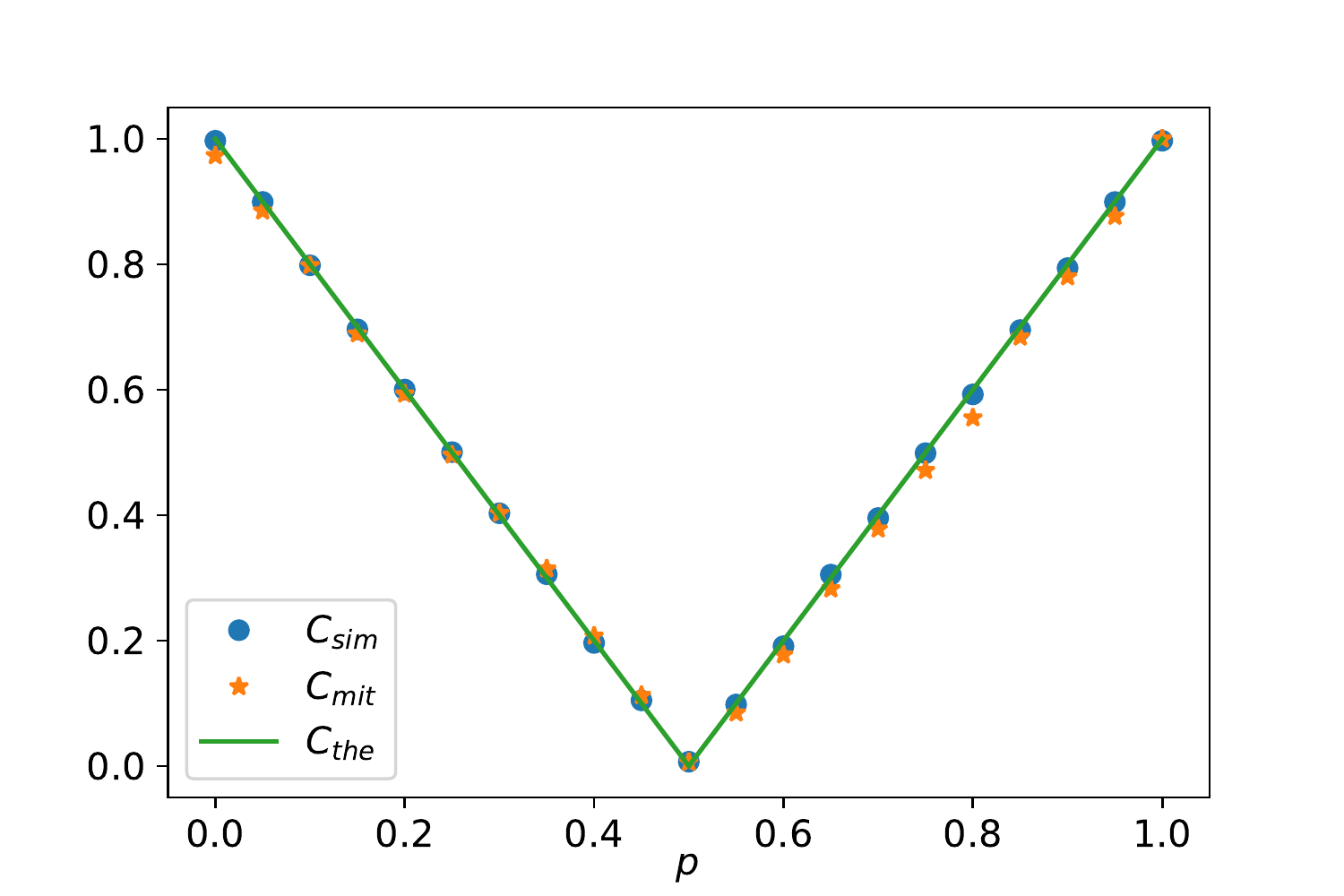}
\caption{Theoretical, simulated, and experimental (with measurement error mitigation) results for the  $l_1$-norm quantum coherence of the state $|+\rangle_A$ under the dynamics generated by the Bit-Phase flip channel implemented using our quantum state preparation-based protocol.}
\label{fig:bpfr}
\end{figure}

As a first specific example, we apply our protocol to simulate the Bit-Phase Flip (BPF) QC model, $\Lambda_{bpf}$, whose 
Kraus' operators are  $K_{00}=\sqrt{1-p}\mathbb{I}$ and $K_{11}=\sqrt{p}Y$ with $p\in[0,1]$.  
With a set of Kraus operators of size two, the ancilla system $B$ can be implemented using one qubit: $|0\rangle_B$ corresponds to the application of $K_{00}$ and $|1\rangle_B$ corresponds to the effect of $K_{11}$ being applied. 
In this particular case, the state to be prepared is
\begin{equation}
|\Psi\rangle_{AB}=\sqrt{\frac{1-p}{2}}|00\rangle-i\sqrt{\frac{p}{2}}|01\rangle+\sqrt{\frac{1-p}{2}}|10\rangle  +i\sqrt{\frac{p}{2}}|11\rangle.
\label{eq:bpf}
\end{equation}
At this point, the protocol is device independent. 
Since we adopted the circuit-based quantum computers from IBMQ, we prepare $|\Psi\rangle_{AB}$ using the algorithm of Ref. \cite{shende} for QSP, that is already implemented in Qiskit \cite{qiskit}. We briefly explain this algorithm in the Appendix.
With this, we present in Fig.~\ref{fig:bpfr} the theoretical ($C_{l_1}(\Lambda_{bpf}(|+\rangle))=|1-2p|$), simulation, and experimental results for the quantum coherence dynamics under the action of the BPF QC.

It is worthwhile mentioning what we mean by theory, simulation, and experiment in the captions of the previous and following figures. By theory we mean that we simply apply the Kraus' representation to obtain the evolved density operator for a given noise channel. By simulation, we mean that we apply our algorithm to simulate a given quantum channel, but that the effect of the quantum gates and measurements are simulated using a classical computer. So, by the simulation we know that our algorithm is working as expected. By its turn, for the experiments we run the quantum circuit of our algorithm, unitary gates and quantum measurements, on real quantum hardware.
Finally, it is worth mentioning that we used the Qiskit tools for measurement error mitigation (QEM) \cite{qiskit} throughout our experimental verification. QEM involves measuring the error syndrome, which is a pattern of errors that occur in the qubits during computation, and using this information to correct the errors. Using QEM improved substantially the experimental results we obtained.

\begin{figure}[t!]
\centering
\includegraphics[width=0.53\textwidth]{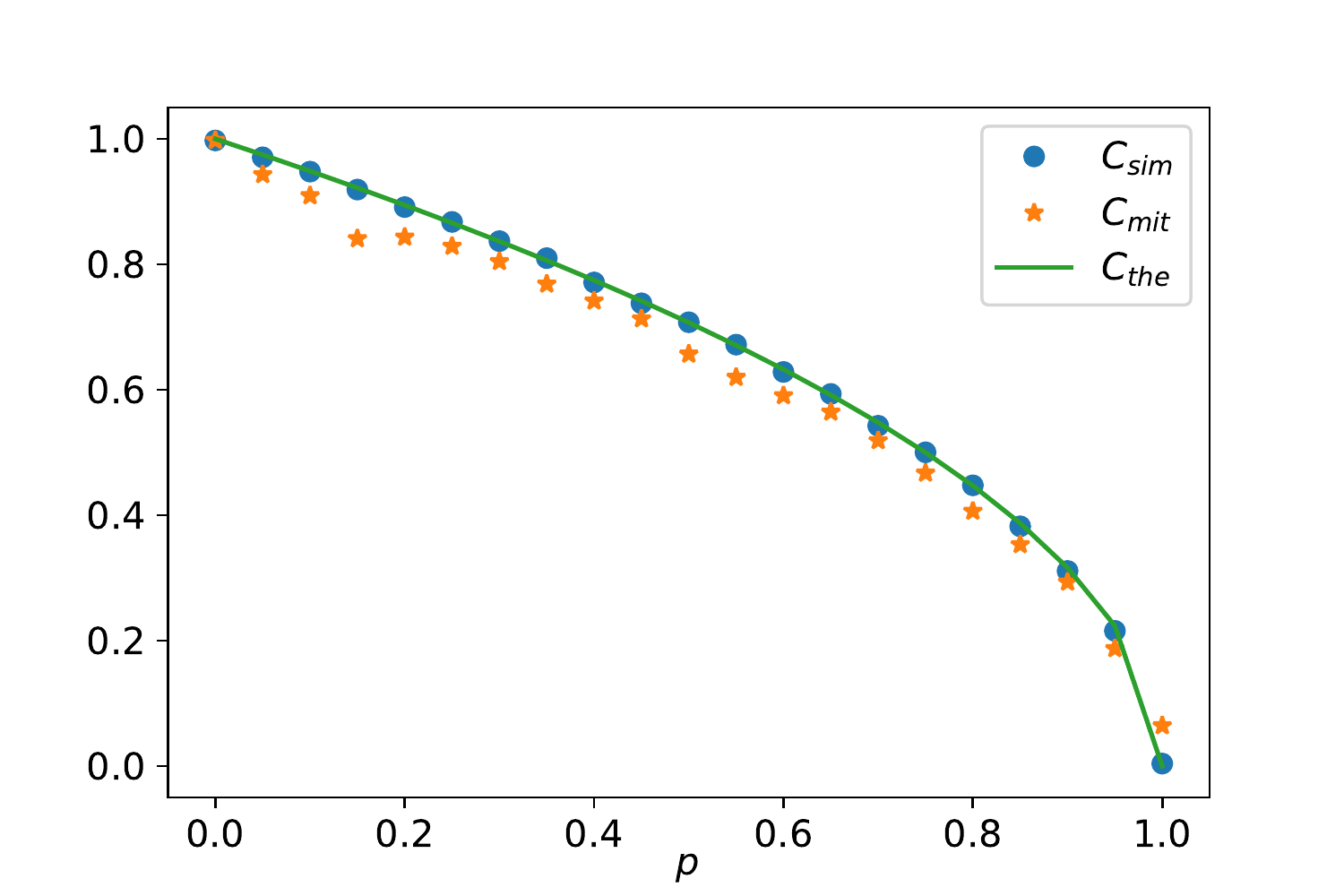}
\caption{Theoretical, simulated, and experimental (with measurement error mitigation) results of the  $l_{1}$-norm quantum coherence of the state $|+\rangle_A$ evolved under the action of the  Phase Damping channel implemented using our quantum state preparation-based protocol.}
\label{fig:pd}
\end{figure}

To conclude this section on one-qubit QC simulation, we execute our protocol for two other important QCs from the literature, namely the Phase Damping (PD), $\Lambda_{pd}$, and Generalized Amplitude Damping (GAD), $\Lambda_{gad}$, QCs, which are related to dephasing and relaxation, respectively \cite{Wilde}.
The PD QC can be described using the following set of Kraus' operators: $K_0=|0\rangle\langle 0|+\sqrt{1-p}|1\rangle\langle 1|$
and  $K_1=\sqrt{p}|1\rangle\langle 1|$,
with $p\in[0,1]$. 
Therefore, as with the BPF QC, the ancilla system $B$ can be implemented with a single qubit. 
Using the same initial state $|\psi\rangle_A=|+\rangle_A$, the state vector to be prepared for the PD QC simulation is 
\begin{equation}    
|\Psi\rangle_{AB}=\frac{1}{\sqrt{2}}|00\rangle+0|01\rangle +\sqrt{\frac{1-p}{2}}|10\rangle+\sqrt{\frac{p}{2}}|11\rangle.
\label{eq:pd}
\end{equation}
The theoretical ($C_{l_1}(\Lambda_{pd}(|+\rangle))=\sqrt{1-p}$), simulation, and experimental results for the dynamics of quantum coherence under the PD QC are shown in Fig. \ref{fig:pd}. 

\begin{figure}[t!]
\centering
\includegraphics[width=0.53\textwidth]{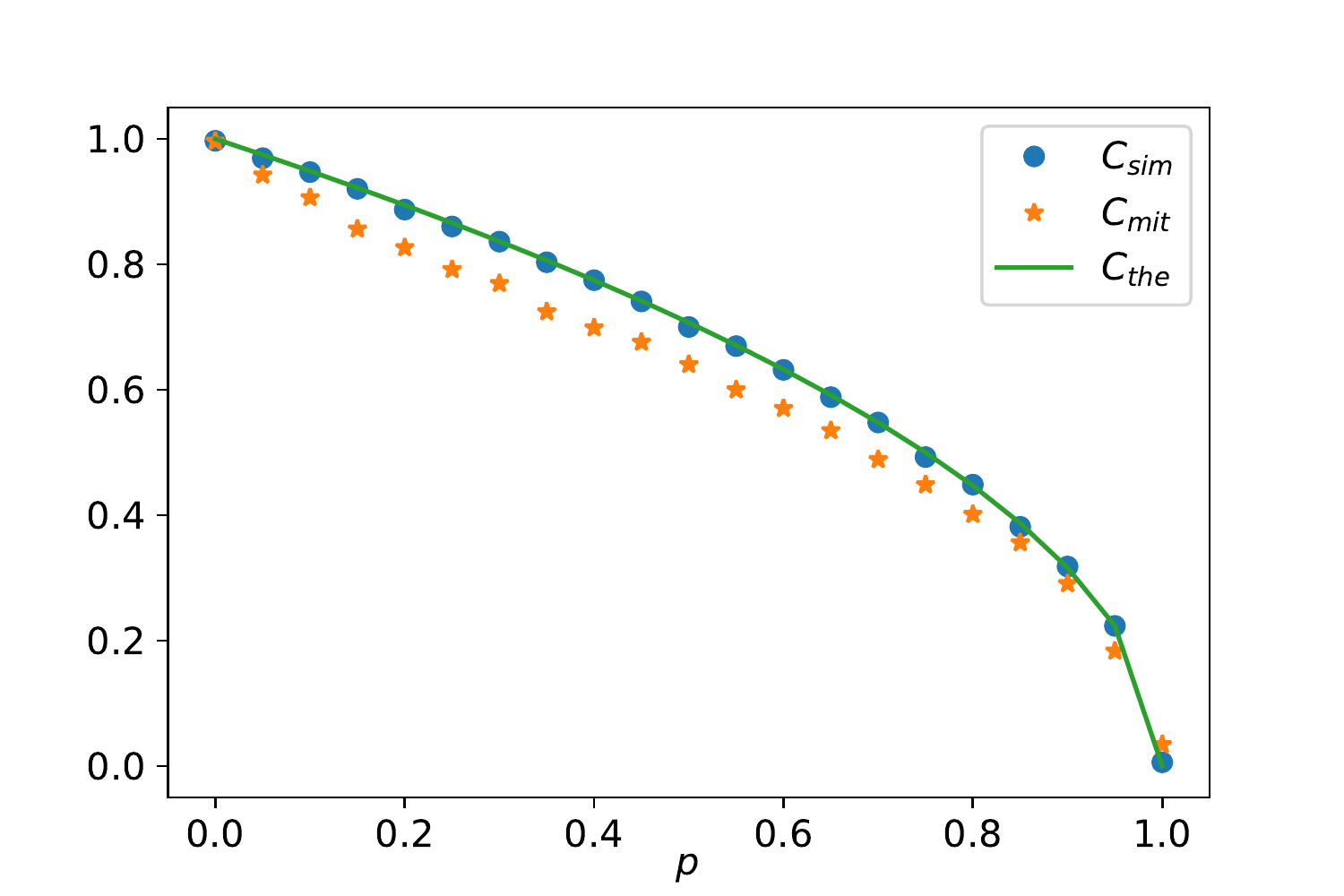}
\caption{Theoretical, simulated, and experimental (with measurement error mitigation) results for the $l_{1}$-norm quantum coherence of the state $|+\rangle_A$ evolved under under the action of the Generalized Amplitude Damping channel implemented using our quantum state preparation-based protocol. For this example, we used $N=0.5$.}
\label{fig:gad}
\end{figure}

Next, let us regard the GAD QC, that is determined by the following set  of Kraus' operators: 
$K_0=\sqrt{1-N}\left(|0\rangle\langle0| +\sqrt{1-p}|1\rangle\langle1|\right)$, $K_1=\sqrt{p\left(1-N\right)}|0\rangle\langle1|$, $K_2=\sqrt{N}\left(\sqrt{1-p}|0\rangle\langle0| +|1\rangle\langle1|\right)$ and $K_3=\sqrt{p N}|1\rangle\langle0|$.
Therefore, the ancilla system $B$ is implemented using two qubits.
It is worthwhile mentioning that two parameters need to be specified for the GAD QC: $p,N\in[0,1]$. 
For the initial state $|\psi\rangle_A=|+\rangle_A$, the state vector to be prepared is
\begin{align} 
\begin{split}
|\Psi\rangle_{AB}&=\sqrt{\frac{1-N}{2}}|000\rangle+ \sqrt{\frac{p\left(1-N\right)}{2}}|001\rangle\\
&+\sqrt{\frac{N(1-p)}{2}}|010\rangle+0|011\rangle\\ &+\sqrt{\frac{(1-N)(1-p)}{2}}|100\rangle+0|101\rangle\\
&+\sqrt{\frac{N}{2}}|110\rangle+\sqrt{\frac{p N}{2}}|111\rangle
\end{split}
\end{align}
Fig.~\ref{fig:gad} presents the theoretical ($C_{l_1}(\Lambda_{gad}(|+\rangle))=\sqrt{1-p}$), simulation, and experimental results for the GAD QC.

For these three one-qubit QCs used for instantiate the application of our QC simulation protocol, we see that the state preparation quantum circuits simulation and experimental execution returned results that are in very good agreement with the theoretical predictions.  

\section{One-qudit noise channels}
\label{sec:qudit}


In this section, we illustrate how to perform the simulation of noisy quantum channels over qudits, $d$-level quantum systems, from the state preparation protocol introduced in the last section. We provide specific examples that involve a qutrit target system under the constraint of the dynamics of the Heisenberg-Weyl dephasing channel or of the amplitude damping channel.

\subsection{Heisenberg-Weyl dephasing channel}

An extension of the Pauli channels for qubits, discussed in the last section, is the Heisenberg-Weyl channel \cite{Wilde}, which can be defined as a probabilistic application, according to a probability distribution $\{p_{j,k}\}$, of the Heisenberg-Weyl operators on a qudit state: 
\begin{align}
\Lambda_{hw}\left(\rho\right)=\sum_{j,k=0}^{d-1}p_{j,k}X\left(j\right)Z\left(k\right)\rho Z\left(k\right)^{\dagger}X\left(j\right)^{\dagger},
\end{align}
with $\sqrt{p_{j,k}}X\left(j\right)Z\left(k\right)=K_{j,k}$ representing the Kraus' operators, where 
$X\left(j\right)=\sum_{k=0}^{d-1}|j\oplus k\rangle\langle k|$ and
$Z\left(j\right)=\sum_{k=0}^{d-1}e^{2\pi ijk/d}|k\rangle\langle k|$  describe the cyclic state shift and phase shift operators, respectively. Above, $\oplus$ is the sum module $d$.
Thus, the global pure quantum state that can be used to simulate such a QC is given by
\begin{align}
& |\Psi\rangle_{AB}=	\sum_{j,k=0}^{d-1}\left(K_{j,k}|\psi\rangle_{A}\right)\otimes|j\rangle_{b}\otimes|k\rangle_{b'} \nonumber \\ 
=&	\sum_{j,k,l=0}^{d-1}\sqrt{p_{j,k}}\psi_{l}e^{2\pi ikl/d}|j\oplus l\rangle\otimes|j\rangle_{b}\otimes|k\rangle_{b'},
\end{align}
where $|\psi\rangle_{A}=\sum_{l=0}^{d-1}\psi_{l}|l\rangle$ is the initial qudit state and $\{|j\rangle\}_s$ is an orthonormal basis for the auxiliary system $s=b,b'$. 
It is worthwhile mentioning that $\Lambda_{hw}$ gives the twirl operation when we consider a uniform probability distribution
$p_{j,k}=1/d^{2}\ \forall j,k$. In this case we obtain $\Lambda_{twirl}\left(\rho\right)=\frac{1}{d^{2}}\sum_{j,k=0}^{d-1}X\left(j\right)Z\left(k\right)\rho X\left(j\right)^{\dagger}Z\left(k\right)^{\dagger}=\mathbb{I}_{d}/d.$

As our first application example, we consider a particular case where the action of the Heisenberg-Weyl channel is reduced to the application of the dephasing channel, which is obtained from $\Lambda_{hw}$ when only the phase shift operators are used:
\begin{equation}
\Lambda_{d}\left(\rho\right)=	\sum_{j=0}^{d-1}p_{j}Z\left(j\right)\rho Z\left(j\right)^{\dagger} 
=	\sum_{k,l=0}^{d-1}\Lambda_{d}\left(\rho\right)_{k,l}|k\rangle\langle l|.
\end{equation}
Such quantum channel does not change the populations, i.e, the diagonal part of the evolved density operator does not change. So, only the coherences are affected by this channel:
$\Lambda_{d}\left(\rho\right)_{k,l}=\rho_{k,l}\left(\sum_{j=0}^{d-1}p_{j}e^{2\pi ij\left(k-l\right)/d}\right)$ for $k\ne l$. 
Let us consider the following probability distribution: $p_{j=0}=p_{0}$ and $p_{0<j\leq d-1}=\left(\frac{1}{d-1}-\frac{p_{0}}{d-1}\right)$.
For the case where the system is a qutrit, $d=3$, the off diagonal elements of the evolved density matrix take the form
\begin{align}
\Lambda_{d}\left(\rho\right)_{k,l}	=\rho_{k,l}\left(p_{0}+p_{1}e^{2\pi i\left(k-l\right)/d}+p_{2}e^{4\pi i\left(k-l\right)/d}\right).
\end{align} 

We consider the $l_1$-norm quantum coherence. For the initial state, we have
$C_{l_{1}}(\rho) 
= 2\big(|\rho_{0,1}|+|\rho_{0,2}|+|\rho_{1,2}|\big).
$
For the evolved state under $\Lambda_{d}$, 
it follows that
$
C_{l_{1}}(\Lambda_{d}(\rho)) 
= 2\big(\big|\Lambda_{d}(\rho)_{0,1}\big|+\big|\Lambda_{d}(\rho)_{0,2}\big|+\big|\Lambda_{d}(\rho)_{1,2}\big|\big)
$
with 
\begin{align}
& \big|\Lambda_{d}(\rho)_{0,1}\big| = \frac{\left|\rho_{0,1}\right|}{2}\sqrt{(p_{0}-p_{1})^{2}+(p_{0}-p_{2})^{2}+(p_{1}-p_{2})^{2}}, \nonumber \\
& \big|\Lambda_{d}(\rho)_{0,2}\big| = \frac{\left|\rho_{0,2}\right|}{2}\sqrt{(p_{0}-p_{1})^{2}+(p_{0}-p_{2})^{2}+(p_{1}-p_{2})^{2}}, \nonumber \\
& \big|\Lambda_{d}(\rho)_{1,2}\big| = \frac{\left|\rho_{1,2}\right|}{2}\sqrt{(p_{0}-p_{1})^{2}+(p_{0}-p_{2})^{2}+(p_{1}-p_{2})^{2}}.
\end{align}

\begin{figure}[t!]
\centering
\includegraphics[width=0.48\textwidth]{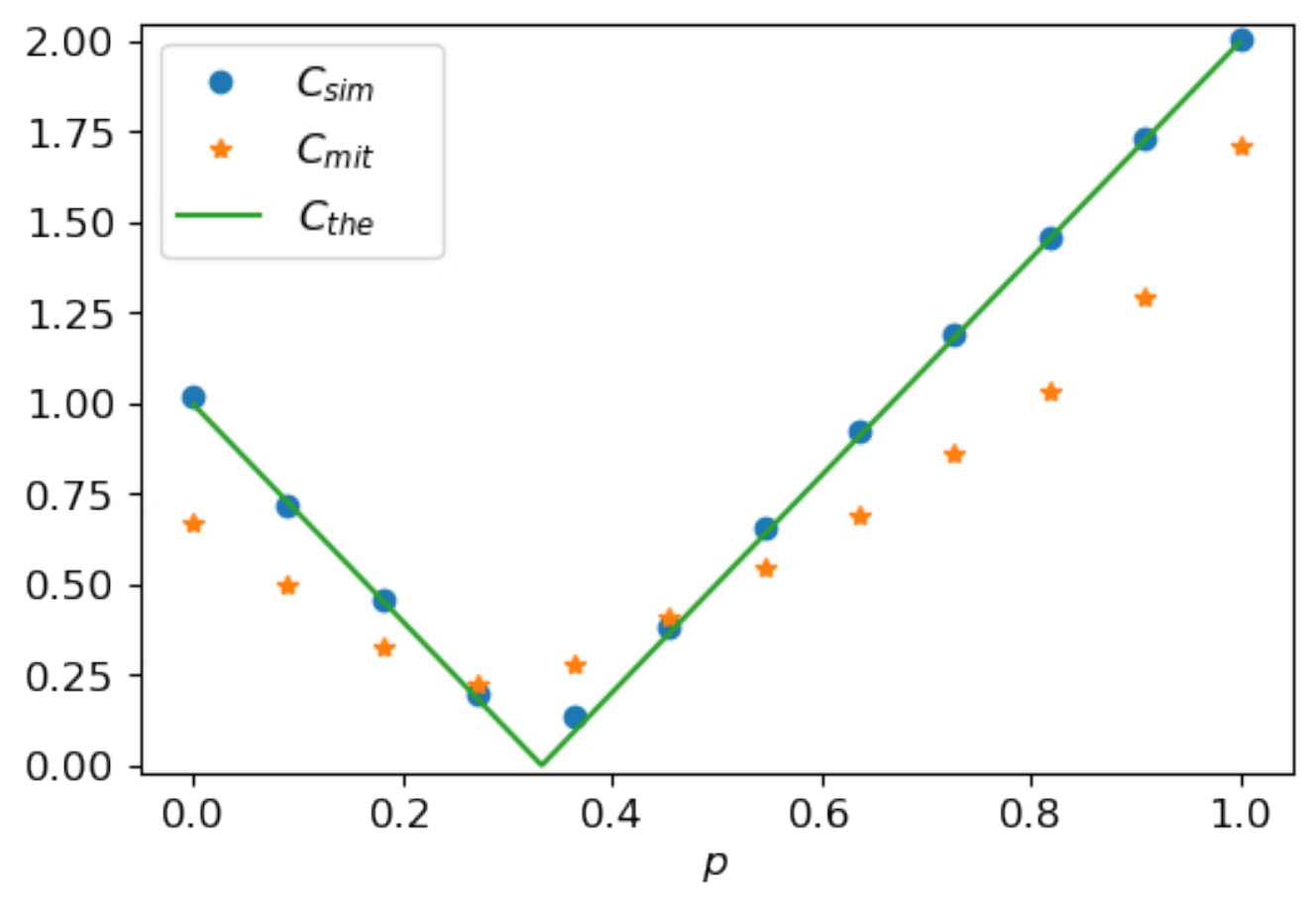}
\caption{Theoretical, simulated and experimental (with measurement error mitigation)  results of the  $l_1$-norm quantum coherence of the state dynamics of one-qutrit under the action of the Heisenberg-Weyl dephasing channel implemented using our quantum state preparation-based  protocol.}
\label{fig:HW}
\end{figure}

In this specific case, considering our protocol, the global state to be prepared for the simulation of the Heisenberg-Weyl dephasing  qutrit channel is given by
\begin{align}
|\Psi_{Aa'bb'}\rangle=&	\sqrt{\frac{p_{0}}{3}}\left(|00\rangle+|01\rangle+|10\rangle\right)_{Aa'}\otimes|00\rangle_{bb'} \nonumber \\
&+\sqrt{\frac{1-p_{0}}{6}}|00\rangle_{Aa'}\otimes|01\rangle_{bb'}  \\&+\sqrt{\frac{\left(1-p_{0}\right)\left(i\sqrt{3}-1\right)^{2}}{24}}|01\rangle_{Aa'}\otimes|01\rangle_{bb'} \nonumber \\	&+\sqrt{\frac{\left(1-p_{0}\right)\left(-i\sqrt{3}-1\right)^{2}}{24}}|10\rangle_{Aa'}\otimes|01\rangle_{bb'} \nonumber \\
&+\sqrt{\frac{1-p_{0}}{6}}|00\rangle_{Aa'}\otimes|10\rangle_{bb'} \nonumber \\
&+\sqrt{\frac{\left(1-p_{0}\right)\left(-i\sqrt{3}-1\right)^{2}}{24}}|01\rangle_{Aa'}\otimes|10\rangle_{bb'} \nonumber \\
&+\sqrt{\frac{\left(1-p_{0}\right)\left(i\sqrt{3}-1\right)^{2}}{24}}|10\rangle_{Aa'}\otimes|10\rangle_{bb'}, \nonumber
\end{align}
where we used the following initial qutrit state $|\psi\rangle_{A}=\big(|0\rangle+|1\rangle+|2\rangle\big)/\sqrt{3}.$ 
The theoretical, simulation, and experimental results for the particular qutrit state regarded here and evolved under Heisenberg-Weyl dephasing is shown in Fig.~\ref{fig:HW}. Once more, we see that our simulation protocol works quite well.

\subsection{Amplitude damping channel}

\begin{figure}[t!]
\centering
\includegraphics[width=0.48\textwidth]{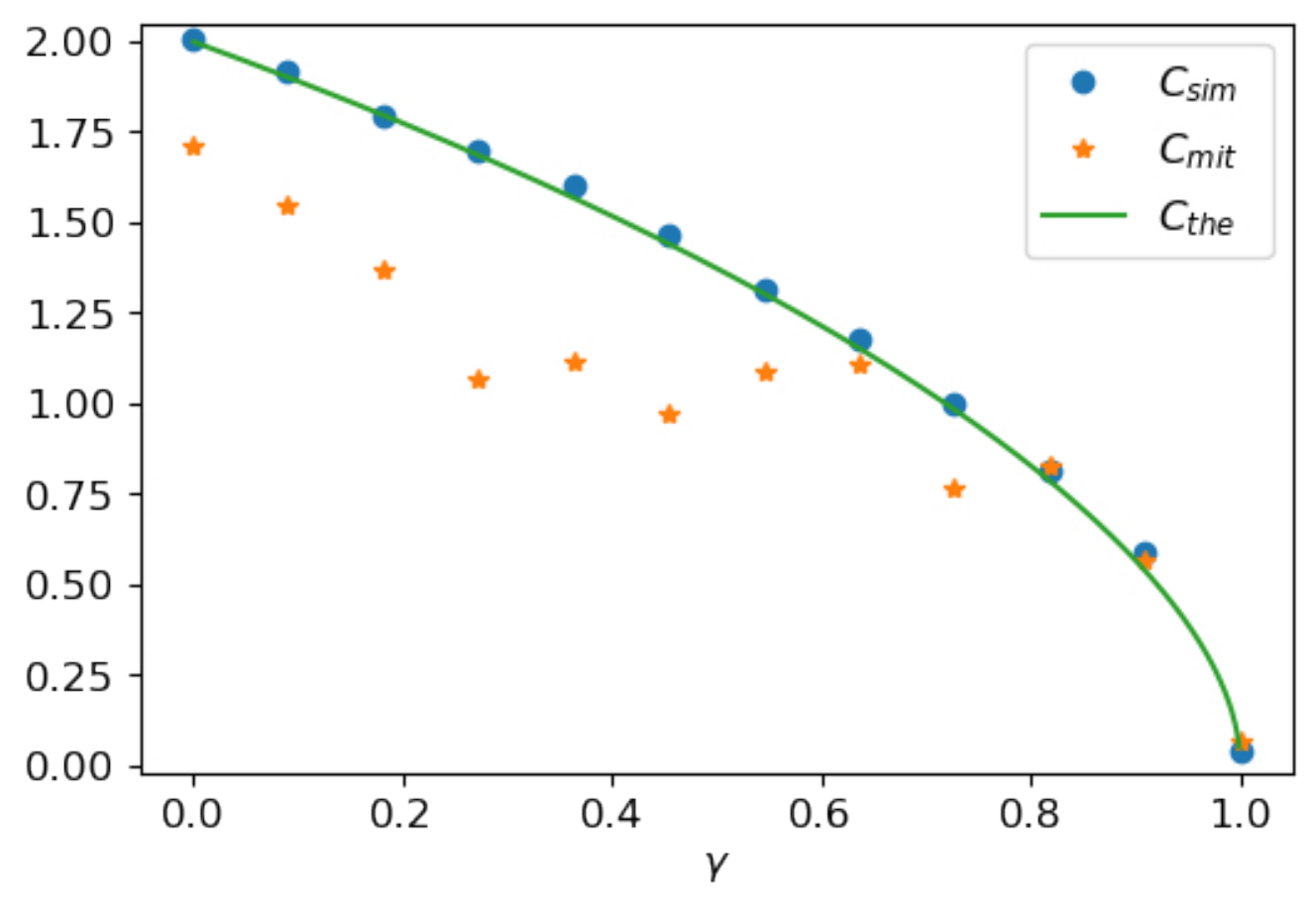}
\caption{Theoretical, simulated and experimental (with measurement error mitigation) results of the  $l_1$-norm quantum coherence of the state dynamics of one-qutrit under the action of the amplitude damping channel implemented using our quantum state preparation-based protocol.}
\label{fig:ad}
\end{figure}

There are qudit noise models that are not directly described by Heisenberg-Weyl-type Kraus' operators and their extensions.
As an example, we consider the  amplitude damping channel (ADC) describing the energy dissipation in a bosonic system at zero temperature.
In this case, the qutrit Kraus' operators 
are given by \cite{Grassl}
\begin{align}
    K_{0}	=&|0\rangle\langle0|+\sqrt{1-\gamma}|1\rangle\langle1|+\left(1-\gamma\right)|2\rangle\langle2|, \\
K_{1}=&\sqrt{\gamma}|0\rangle\langle1|+\sqrt{2\gamma\left(1-\gamma\right)}|1\rangle\langle2|, \\
K_{2}	=&\gamma|0\rangle\langle2|.
\end{align}
So, the ADC,
$\Lambda_{ad}\left(\rho\right)=\sum_{j=0}^{2}K_{j}\rho K_{j}^{\dagger}
$,
can be simulated by the state preparation method by producing the state
\begin{align}
    & |\Psi\rangle_{Aa'bb'}
=	\frac{1}{\sqrt{3}}|0000\rangle_{Aa'bb'}+\sqrt{\frac{1-\gamma}{3}}|0100\rangle_{Aa'bb'} \nonumber  \\
	&+\frac{1-\gamma}{\sqrt{3}}|1000\rangle_{Aa'bb'}+\frac{\gamma}{\sqrt{3}}|0010\rangle_{Aa'bb'} \\
&+\sqrt{\frac{\gamma}{3}}|0001\rangle_{Aa'bb'}+\sqrt{\frac{2\gamma\left(1-\gamma\right)}{3}}|0101\rangle_{Aa'bb'},\nonumber
\end{align}
with the initial qutrit state being set to $|\psi\rangle_{A}=\big(|0\rangle+|1\rangle+|2\rangle\big)/\sqrt{3}.$
The theoretical,
 $   C_{l_{1}}\left(\Lambda_{ad}\left(\rho\right)\right)=	\frac{2}{3}\left(\left|\left(1-\gamma\right)^{\frac{3}{2}}\right|+\left|1-\gamma\right|\right) 
	+\frac{4}{3}\left|\left(\sqrt{2}\gamma+1\right)\sqrt{1-\gamma}\right|,
$
simulation, and experimental results for $l_1$-norm coherence of this qutrit state under the ADC are shown in Fig. \ref{fig:ad}. The simulation results match the theory, and the experimental data agree fairly well with the theoretical prediction. Here we also applied measurement error mitigation, directly implemented through Qiskit \cite{qiskit}, what improved considerably the experimental results.

\section{Lorentz transformations}
\label{sec:lorentz}

In this section, we discuss how to simulate Lorentz transformations of spin density matrices (i.e. the Wigner rotations) for a massive spin-$1/2$ quantum particle in a circuit-based quantum computer. To do this, we remember that a Lorentz boost induces a transformation on the spin states that depends on the momentum states of the particle \cite{Terno}. Therefore, the unitary representation of the Lorentz transformations can be seen as a controlled quantum operation (or a controlled-$U$ gate) where the momentum works as the control system, while the spin plays the role of a target qubit \cite{Palge}. Besides, in this work, we restrict ourselves to discrete momentum states once the momentum states will be implemented using qubits in a circuit-based quantum computer simulation protocol. 

Let us now suppose that an inertial observer $\mathcal{O}$ describes the state of a spin-$1/2$ particle as
\begin{align}
    |\Psi\rangle = \sum_{p, \lambda} \psi_{\lambda}(p) |p,\lambda\rangle,
\end{align}
where $|p,\lambda\rangle := |p\rangle \otimes |\lambda\rangle$ is a basis vector in the composite Hilbert space $\mathcal{H}_p \otimes \mathcal{H}_s$ where $p$ labels the momentum state and $\lambda$ the spin state of the particle. Given another inertial observer $\mathcal{O}'$, that is related to $\mathcal{O}$ through a Lorentz boost $\Lambda$, it is well known that $\mathcal{O}'$ assigns a different state $|\Psi_{\Lambda}\rangle := U(\Lambda) |\Psi\rangle$ to the same quantum particle, where $U(\Lambda)$ is a unitary representation of the Lorentz boost $\Lambda$ \cite{Weinberg}. The action of $U(\Lambda)$ is defined by \cite{Onuki}:
\begin{align}
    U(\Lambda) |p, \lambda\rangle := |\Lambda p\rangle \otimes D(W(\Lambda,p)) |\lambda\rangle,
\end{align}
where $W(\Lambda,p)$ is called a Wigner rotation. The set of Wigner rotations forms a group known as the Wigner's little group (WLG), which is a subgroup of the Poincar\'e group. Therefore, one can see that, under a general Lorentz boost $\Lambda$, the momenta $p$ goes to $\Lambda p$ and the spin transforms under unitary controlled operation $D(\Lambda,p)$, which is a representation of an element $W(\Lambda,p)$ of the WLG. For massive quantum particles, it is known that the WLG is the group of rotations in three dimensions, $SO(3)$, which, in turn, is homomorphic to the special unitary group $SU(2)$ \cite{Tung}. Since we are dealing with spin-$1/2$ particles, which embodies the notion of a qubit, the unitary representation of the Wigner rotation is given by \cite{Ahn, Halpern}
\begin{align}
    D(W(\Lambda, p)) =  \cos \frac{\theta}{2} \mathbb{I} + i \sin \frac{\theta}{2} (\vec{\sigma} \cdot \hat{n}), \label{eq:wigrot}
\end{align}
with $\mathbb{I}$ being the identity matrix, $\vec{\sigma}=(\sigma_x,\sigma_y,\sigma_z)$ is a vector whose components are the Pauli matrices, and
\begin{align}
    & \cos \frac{\theta}{2} = \frac{\cosh\frac{\omega}{2}\cosh\frac{\alpha}{2} + \sinh\frac{\omega}{2}\sinh\frac{\alpha}{2} (\hat{e} \cdot \hat{p})}{\sqrt{\frac{1}{2}(1 + \cosh \omega \cosh \alpha + \sinh \omega \sinh \alpha (\hat{e} \cdot \hat{p}))}},\\
    & \sin \frac{\theta}{2} \hat{n} = \frac{ \sinh\frac{\omega}{2}\sinh\frac{\alpha}{2} (\hat{e} \times \hat{p})}{\sqrt{\frac{1}{2}(1 + \cosh \omega \cosh \alpha + \sinh \omega \sinh \alpha (\hat{e} \cdot \hat{p}))}},
\end{align}
where $\cosh \alpha = p^0/m$, $\omega = \tanh^{-1} v$ is the rapidity of the boost, $\hat{e}$ is the unit vector pointing in the direction of the boost, $p$ is the $4$-momenta of the particle in $\mathcal{O}$, and $\Lambda p$ is the $4$-momenta of the particle in $\mathcal{O}'$. Reinforcing, it is worth to notice that the Lorentz boost can be thought as a controlled unitary. Given that the boost angle and rapidity are fixed, then the spin transformation depends solely on the momentum state \cite{Palge}, i.e.
\begin{align}
    U(\Lambda) = \sum_j |\Lambda p_j\rangle\langle p_j| \otimes D(W(\Lambda, p_j)).
\end{align}

Now, let us describe how to put the Wigner's rotations in terms of the maps of the kind used in open quantum dynamics, i.e., in the form of the Kraus' operator sum representation. Given an orthonormal basis of momentum states $\{ |p_j\rangle \}_{j = 0}^{d_p - 1}$, where $d_p = \dim \mathcal{H}_p$, and the following separable state described by $\mathcal{O}$
\begin{align}
    |\Psi\rangle = \frac{1}{\sqrt{d_p}} \sum_{j = 0}^{d_p -1} |p_j\rangle \otimes |\psi\rangle, \label{eq:psi}
\end{align}
where $|\psi\rangle \in \mathcal{H}_s$, in $\mathcal{O}'$, the global state of the system is described by
\begin{align}
    |\Psi_{\Lambda}\rangle = \frac{1}{\sqrt{d_p}} \sum_{j = 0}^{d_p -1} |\Lambda p_j\rangle \otimes D(W(\Lambda,p_j)) |\psi\rangle,
\end{align}
which, in general, is an entangled state. Meanwhile, the reduced spin density matrix in $\mathcal{O}'$ is given by
\begin{align}
    \Lambda(\rho_s) & := \rho_{\Lambda s} := \Tr_p (\ketbra{\Psi_{\Lambda}}) \nonumber \\
    & = \sum_{j = 0}^{d_p -1} \Big(\frac{1}{\sqrt{d_p}}  D(W(\Lambda,p_j))\Big) \rho_s \Big(\frac{1}{\sqrt{d_p}}  D^{\dagger}(W(\Lambda,p_j))\Big) \nonumber \\
    & = \sum_{j = 0}^{d_p -1} K_j \rho_s K^{\dagger}_j, \label{eq:lambs}
\end{align}
where $\rho_s := |\psi\rangle\langle\psi| = \Tr_p (\ketbra{\Psi})$ with $|\Psi\rangle$ expressed by Eq. (\ref{eq:psi}) and $K_j := \frac{1}{\sqrt{d_p}}  D(W(\Lambda,p_j))$. By Eq. (\ref{eq:wigrot}), it is straightforward to see that
\begin{align}
    \sum_{j = 0}^{d_p -1} K^{\dagger}_j K_j = \frac{1}{d_p} \sum_{j = 0}^{d_p -1} D^{\dagger}(W(\Lambda,p_j)) D(W(\Lambda,p_j)) = \mathbb{I}.  
\end{align}
Besides, one can see that $ \sum_{j = 0}^{d_p -1} K_j K_j^{\dagger} = \mathbb{I}$ also holds, which implies that the map is unital and the identity is preserved, $\Lambda(\mathbb{I}) = \mathbb{I}$. For the initial density matrices of the type $\rho_s = \ketbra{\psi}$, it is easy to see that the map $\Lambda(\rho_s) := \rho_{\Lambda s}$ defined by Eq. (\ref{eq:lambs}) is positive, once for any $\ket{\phi} \in \mathcal{H}_s $ and $\rho_s = \ketbra{\psi}$ we have
\begin{align}
    \expval{\Lambda(\rho_s)}{\phi} = \frac{1}{d_p} \sum_{j = 0}^{d_p -1} \abs{\bra{\phi}D(W(\Lambda,p_j)) \ket{\psi}}^2 \ge 0.
\end{align}
The subject of completely positivity of the map $\Lambda(\rho_s) = \rho_{\Lambda s}$ is more subtle. As already noticed in Ref. \cite{Jordan}, the spin and momentum states can be initially entangled. Therefore the map $\Lambda(\rho_s)$ is generally non completely positive and acts in limited domains as described in Ref. \cite{Shaji}. However, for initially product states between momentum and spin, as the one considered in the Eq. (\ref{eq:psi}), it is straightforward to see that $\Lambda(\rho_s)$ is completely positive, once $\bra{\Phi} \Lambda \otimes \mathbb{I}(\ketbra{\Psi}) \ket{\Phi} \ge 0 \ \forall \ket{\Phi} \in \mathcal{H}_p \otimes \mathcal{H}_s$. Therefore, for the class of initial separable states considered in this work, the map $\Lambda$ is CPTP. As for initially entangled states of the type $ \ket{\Upsilon} = \frac{1}{\sqrt{d_p}} \sum_{j = 0}^{d_p -1} \ket{p_j} \otimes \ket{\psi_j}$, we notice that $\rho_{\Lambda s} = \sum_{j = 0}^{d_p -1} K_j \rho^s_j K^{\dagger}_j$ where $\rho^s_j = \ketbra{\psi_j}$ and therefore we do not have a Kraus-like map of the form given by Eq. (\ref{eq:lambs}). Finally, it is worth mentioning that the limit on the number of Kraus operators needed to describe a quantum dynamics is $d_s^2$, where $d_s = \dim \mathcal{H}_s$. In our case, $d_s = 2$. As one can see, the number of Kraus operators for the Wigner rotations, in principle, depends on the dimension of Hilbert space for the momentum. However, it is well known that the Kraus operators are not unique and sets of operators $\{ K'_j\}$ and $\{K_j\}$ that are related by a unitary transformation generate the same quantum dynamics. Therefore, the map generated by Eq. (\ref{eq:lambs}) for spin-$1/2$ particles can, in principle, be generated by only four Kraus operators.

\begin{figure}[t]
\centering
\includegraphics[width=0.48\textwidth]{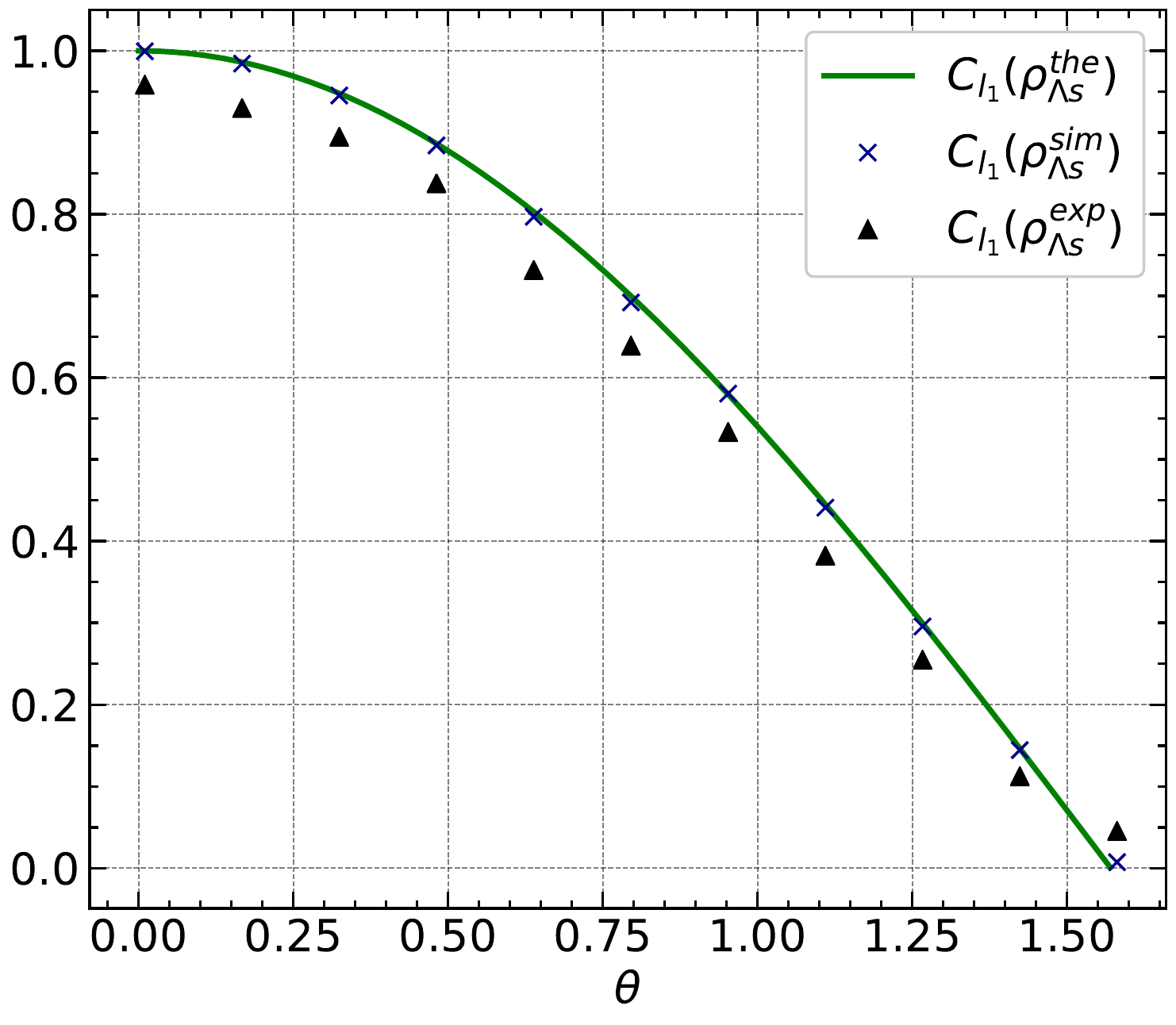}
\caption{Theoretical, simulated, and experimental results of the  $l_{1}-$norm quantum coherence of the spin state given by Eq.~(\ref{eq:spinrot}) under the action of the Wigner rotations implemented using the quantum state preparation-based protocol.}
\label{fig:lorentz}
\end{figure}

Lastly, let us consider an specific example to implement the simulation of the Wigner rotations using our protocol. If the momentum is in the $x$-direction of the reference frame $\mathcal{O}$, such that initial state is given by
\begin{align}
    \ket{\Psi} = \frac{1}{2}(\ket{p} + \ket{-p}) \otimes (\ket{\uparrow} + \ket{\downarrow}),
\end{align}
where the quantization spin axis is along the $z$-direction. Therefore, both momentum and spin degrees of freedom can be described as qubits. Besides, let us consider that the boost is given in the $z$-axis. Then $D(W(\Lambda, p)) =  \cos \frac{\theta}{2} \mathbb{I} + i \sin \frac{\theta}{2} \sigma_y$. Therefore, the state described by $\mathcal{O'}$, to be prepared in the circuit-based quantum computer using state preparation algorithms, is given by
\begin{align}
    & 2\ket{\Psi_{\Lambda}} \\
    & = \ket{\Lambda p} \otimes \Big((\cos \frac{\theta}{2} + \sin \frac{\theta}{2})\ket{\uparrow} + (\cos \frac{\theta}{2} - \sin \frac{\theta}{2}) \ket{\downarrow} \Big) \nonumber \\
    & + \ket{-\Lambda p} \otimes \Big((\cos \frac{\theta}{2} - \sin \frac{\theta}{2})\ket{\uparrow} + (\cos \frac{\theta}{2} + \sin \frac{\theta}{2}) \ket{\downarrow} \Big), \nonumber
\end{align}
with $\theta \in [0, \pi/2]$. The corresponding Kraus operators are given by
\begin{align}
  K_0=\frac{1}{\sqrt{2}} \begin{bmatrix} \cos \frac{\theta}{2} & \sin \frac{\theta}{2}\\ - \sin \frac{\theta}{2} & \cos \frac{\theta}{2}  \\ \end{bmatrix}, \ \ K_1 = \frac{1}{\sqrt{2}} \begin{bmatrix} \cos \frac{\theta}{2} & - \sin \frac{\theta}{2}\\ \sin \frac{\theta}{2} & \cos \frac{\theta}{2}  \\ \end{bmatrix}.   \nonumber
\end{align}
The reduced spin density matrix is
\begin{align}
    \rho_{\Lambda s} = \frac{1}{2} \begin{bmatrix} 1 & \cos \theta \\ \cos \theta & 1  \\ \end{bmatrix}. \label{eq:spinrot}
\end{align}
For illustrating the application of our protocol, we prepare this state $\ket{\Psi_{\Lambda}}$ for some values of $\theta$. After performing state tomography and taking the partial trace over the momentum degree of freedom, we compute the quantum coherence of the spin state. The results are shown in Fig. \ref{fig:lorentz}. The simulated results match the theory and the experimental results agree fairly well with the theoretical predictions.


\section{Mixed initial states}
\label{sec:mixed}

In the previous sections, we discussed how to simulate noisy quantum channels (NQCs) for the system prepared initially in a pure state. In this section we identify three ways we can use this knowledge to simulate NQCs applied to mixed initial states.

First, we notice that if we know $\rho_{A}=\sum_{l,m=0}^{d-1}\rho_{l,m}^A|l\rangle_A \langle m|$ and $\{K_{j}\}$, we can use Eq.~(\ref{eq:iso}) to compute 
\begin{align}
& \tilde{\rho}_{AB} = V_{AB}(\rho_A \otimes|0\rangle_B \langle 0|V_{AB}^\dagger\big) \\
 & = \sum_{l,m=0}^{d-1}\rho_{l,m}^{A}\sum_{j}\big(K_{j}|l\rangle_{A}\big)\otimes|j\rangle_{B}\sum_{k}\big(\langle m|_{A}K_{k}^{\dagger}\big)\otimes\langle k|_{B}.  \nonumber 
\end{align}
Once we computed and diagonalized this density matrix, $\tilde{\rho}_{AB} = \sum_{j=0}^{d_{AB}-1}\lambda_{j}|\lambda_{j}\rangle_{AB}\langle \lambda_{j}|,$ we can obtain the evolved state under the quantum operation $\Lambda$ by preparing the mixed state $\tilde{\rho}_{AB}$ through the purification
$|\Phi\rangle_{ABC} = \sum_{j=0}^{d_{AB}-1}\sqrt{\lambda_{j}}|\lambda_{j}\rangle_{AB}\otimes|j\rangle_{C}$.
By taking the partial trace over $B$ and $C$, we finally obtain
$\Lambda(\rho_{A}) = \Tr_{BC}\big(|\Phi\rangle_{ABC}\langle\Phi|\big)$.

In a second way to simulate NQCs applied to initially mixed states, we start noticing that any density matrix can be diagonalized, $\rho_{A} = \sum_{k=0}^{d_{A}-1}r_{k}|r_{k}\rangle_{A}\langle r_{k}|$,
and we use the linearity of the NQC: $\Lambda(\rho_{A}) =\sum_{k=0}^{d_{A}-1}r_{k}\Lambda(|r_{k}\rangle).$
So, we first obtain the spectral decomposition of $\rho_{A}$, then we prepare the states
$|\Psi_{k}\rangle_{AB} = \sum_{j}\big(K_{j}|r_{k}\rangle\big)\otimes|j\rangle_{B}$
for
$k=0,\cdots,d_{A}-1$
and we trace over the subsystem $B$ to obtain 
$\Lambda(|r_{k}\rangle) = \Tr_{B}\big(|\Psi_{k}\rangle_{AB}\langle\Psi_{k}|\big)$.
Finally, we compute the statistical mixture 
$\sum_{k}r_{k}\Lambda(|r_{k}\rangle)=\Lambda(\rho_A).$

For obtaining a third method for simulating quantum channels applied to mixed initial states, we start by considering a purification $|\Psi\rangle_{AB} = \sum_j \sqrt{r_j }|r_j\rangle_A \otimes|j\rangle_B$ of system $A$ state $\rho_A = \sum_j r_j |r_j\rangle\langle r_j|$. One can verify that $\Tr_B \big(\Lambda_A \otimes\mathbb{I}_B(|\Psi\rangle_{AB})\big) = \Tr_B \big(\sum_j K_j \otimes\mathbb{I}_B |\Psi\rangle_{AB}\langle\Psi|K_j^\dagger \otimes\mathbb{I}_B\big) = \Lambda(\rho_A)$. Now, analogously to the pure initial state case, we  regard the isometry 
$V_{ABC}|\Psi\rangle_{AB}\otimes|0\rangle_C = \sum_j K_j\otimes\mathbb{I}_B |\Psi\rangle_{AB} \otimes|j\rangle_C =:|\Phi\rangle_{ABC}$, from which we obtain $\Tr_C \big(V_{ABC}|\Psi\rangle_{AB}\otimes|0\rangle_C \langle\Psi|_{AB} \otimes\langle 0|_C V_{ABC}^\dagger \big) = \Lambda_A \otimes\mathbb{I}_B(|\Psi\rangle_{AB})$. With this, we see that the quantum channel $\Lambda(\rho_A) = \Tr_{BC}\big(|\Phi\rangle_{ABC}\langle\Phi|\big)$
can be simulated by preparing the pure quantum state
\begin{equation}
|\Phi\rangle_{ABC} = \sum_{j,l} \sqrt{r_l}K_j|r_l\rangle_A \otimes|l\rangle_B \otimes |j\rangle_C
\end{equation}
and taking the partial trace over the auxiliary systems $B$ and $C$. We observe that the dimension of $B$ is equal to the rank of $\rho_A$ and the dimension of $C$ is equal to the number of Kraus' operators used to represent the quantum channel $\Lambda$.

As an application example, let us use this third method, which requires less auxiliary qubits and classical processing, to simulate the depolarizing channel 
\begin{equation}
    \Lambda_d (\rho_A) = (1-p)\rho_A + p\mathbb{I}/2
\end{equation}
applied to a mixed qubit state. This quantum channel is obtained from the Pauli channel in Sec. \ref{sec:qubit} by setting
\begin{equation}
p_I = (4-3p)/4,\ p_X = p_Z = p_Y = p/4. \label{eq:pr_dep}
\end{equation}
Using our method, one can simulate any Pauli channel by preparing the state
\begin{align}
& |\Phi\rangle_{ABCD} = \label{eq:psi_dep} \\
& \sqrt{r_0 p_I}\cos(\theta/2)|0000\rangle + \sqrt{r_0 p_I}e^{i\phi}\sin(\theta/2)|1000\rangle \nonumber \\
& + \sqrt{r_0 p_X}\cos(\theta/2)|1001\rangle + \sqrt{r_0 p_X}e^{i\phi}\sin(\theta/2)|0001\rangle \nonumber \\
& + \sqrt{r_0 p_Z}\cos(\theta/2)|0010\rangle - \sqrt{r_0 p_Z}e^{i\phi}\sin(\theta/2)|1010\rangle \nonumber \\
& + i\sqrt{r_0 p_Y}\cos(\theta/2)|1011\rangle - i\sqrt{r_0 p_Y}e^{i\phi}\sin(\theta/2)|0011\rangle \nonumber \\
& + \sqrt{r_1 p_I}\sin(\theta/2)|0100\rangle - \sqrt{r_1 p_I}e^{i\phi}\cos(\theta/2)|1100\rangle \nonumber \\
& + \sqrt{r_1 p_X}\sin(\theta/2)|1101\rangle - \sqrt{r_1 p_X}e^{i\phi}\cos(\theta/2)|0101\rangle \nonumber \\
& + \sqrt{r_1 p_Z}\sin(\theta/2)|0110\rangle + \sqrt{r_1 p_Z}e^{i\phi}\cos(\theta/2)|1110\rangle \nonumber \\
& + i\sqrt{r_1 p_Y}\sin(\theta/2)|1111\rangle + i\sqrt{r_1 p_Y}e^{i\phi}\cos(\theta/2)|0111\rangle, \nonumber 
\end{align}
and tracing over the auxiliary qubits $B,C,D$. For simulating $\Lambda_d$, we use the probabilities given in Eq. (\ref{eq:pr_dep}). 
\begin{figure}[t]
\centering
\includegraphics[width=0.49\textwidth]{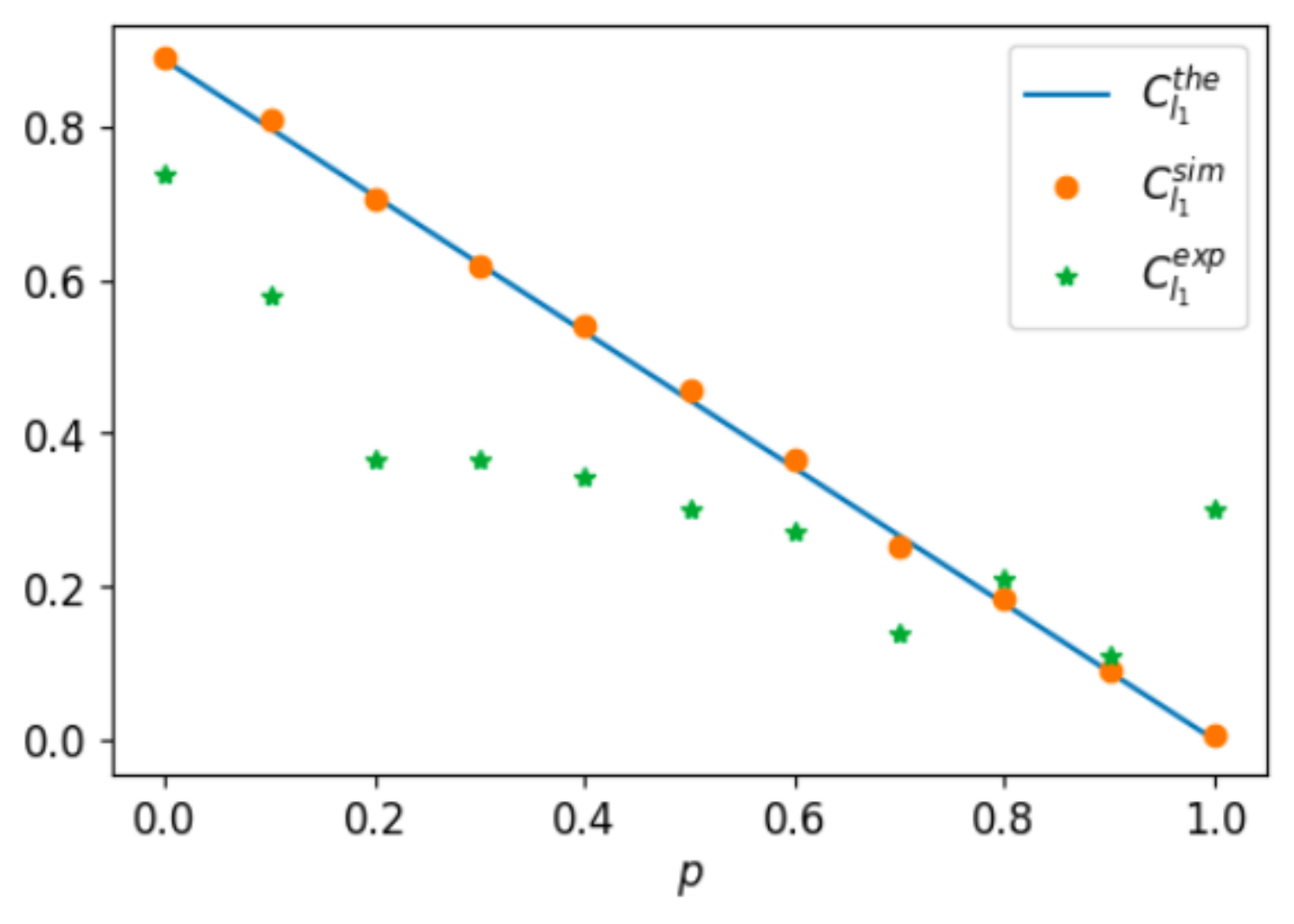}
\caption{Theoretical, simulated, and experimental results of the  $l_{1}$-norm quantum coherence for the mixed initial state of Eq. (\ref{eq:mixed}) evolved under the action of the depolarizing quantum channel  implemented using the quantum state preparation-based protocol.}
\label{fig:mixed}
\end{figure}
In Eq. (\ref{eq:psi_dep}), we have 
\begin{align}
& r = \sqrt{(\rho_{00}-\rho_{11})^2 + |2\rho_{01}|^2}, \\
& \theta = \arccos\frac{\rho_{00}-\rho_{11}}{r}, \\
& \phi = \arccos \frac{\Re(2\rho_{01})}{r\sin\theta},
\label{eq:mixed}
\end{align}
with $\rho_{j,k}$ being the matrix elements of $\rho_A$ when represented in the computational basis. Besides, $r_j = \frac{1+(-1)^j r}{2}$, for $j=0,1$, are the eigenvalues of $\rho_A$. Even though by preparing this state one can simulate $\Lambda_d$ applied to any initial state, let us consider the following specific example
\begin{equation}
\rho_A = \begin{bmatrix}2/3&1.33/3 \\ 1.33/3&1/3\end{bmatrix}.
\end{equation}
For this initial state, we show in Fig.~\ref{fig:mixed} the theoretical, simulation, and experimental results for the dynamics of $l_1$-norm coherence under the depolarizing channel. For these experiments, we used the $ibmq\_ 
quito$ quantum chip with the following correspondence for the quantum registers: $\text{qr}[1]=A$, $\text{qr}[0]=D$, $\text{qr}[3]=C$, and $\text{qr}[2]=B$.

\section{Final remarks}
\label{sec:conc}

Having a good understanding of noise quantum channels (NQCs) is of uppermost importance for the future of quantum technologies. And the simulation of NQCs is a fundamental tool for acquiring this knowledge. In the last few years, quantum state preparation algorithms (QSPAs) have gained a distinguished role in quantum computing and quantum simulation. In this article, we showed that one can avoid the practical difficulties involved in some NQCs simulation techniques from the literature. We showed that NQCs can be simulated using QSPAs and we instantiated the ease of use of our technique applying it to examples of qubit and qudit NQCs and for spin NQCs originating from Lorentz transformations. We also investigated the simulations of NQCs applied to mixed initial states. Most of these examples were implemented experimentally using IBM's quantum processors. While the classical simulation matched almost exactly the theory, even with the high noise rate of the used quantum computers, our experimental results agreed fairly well with the theoretical predictions. So, we believe that our protocol will facilitate the practical use of NQCs simulations, fostering thus further related research in quantum information science. 

Finally we must recognize the great amount of work that researches have being doing lately regarding quantum state preparation (QSP), that has a central role in quantum computation, quantum simulation, and quantum machine learning. Besides, although one already knows that the complexity for preparing a general unknown quantum state scales exponentially with the number of qubits, we also know that if we restrict in some way the set of states to be prepared, much more efficient algorithms can be produced. Besides the more immediate advantages of our algorithm we mentioned above, it is in the light of this observation that comes about one of the main future prospects for advantage of our quantum channel simulation algorithm related to others algorithms from the literature. This is so because our algorithm paves the way for the discovery of more efficient quantum channel simulation algorithms via the restrictions imposed by the quantum channel symmetries on the possible set of generated quantum states, and on the developments made in QSP algorithms research. Besides, a promising tool to be used in this direction is non-deterministic QSP algorithms, as for instance the variational quantum state preparation algorithm.

\begin{acknowledgments}
This work was supported by the S\~ao Paulo Research Foundation (FAPESP), Grant No.~2022/09496-8.3, by the National Institute for the Science and Technology of Quantum Information (INCT-IQ), process 465469/2014-0, by the Coordination for the Improvement of Higher Education Personnel (CAPES), process 88882.427913/2019-01, by the National Council for Scientific and Technological Development (CNPq), process 309862/2021-3, and by the Brazilian Space Agency (AEB), process 01350.001732/2020-61 (TED 020/2020).
\end{acknowledgments}

\vspace{0.3cm}
\textbf{Data availability.}
The data that support the findings of this study are available upon reasonable request from the authors.


\appendix*
\section{The quantum state preparation algorithm utilized}
\label{sec:ap}

In this appendix, we briefly explain the algorithm we use in this work for quantum state preparation. It is based on Ref. \cite{shende}. 

For stating the main idea of the algorithm, let us start considering a one-qubit state:
\begin{align}
|\psi\rangle & = |c_{0}|e^{i\phi_{0}}|0\rangle + |c_{1}|e^{i\phi_{1}}|1\rangle \\
& = e^{it/2}\big(e^{-i\phi/2}\cos(\theta/2)|0\rangle + e^{i\phi/2}\sin(\theta/2)|1\rangle\big),\nonumber
\end{align}
with $\phi=\phi_{1}-\phi_{0}$, $t = \phi_{1}+\phi_{0}$, and $\theta = 2\arccos(|c_0|)\in[0,\pi].$ Regarding the representation of this state on the Bloch sphere \cite{Nielsen}, it is not difficult to see how to lead this state to the $z$ axis via the rotations $R_{z}(\phi) = \begin{bmatrix} e^{-i\phi/2} & 0 \\ 0 & e^{i\phi/2} \end{bmatrix}$ and $R_{y}(\theta) = \begin{bmatrix} \cos(\theta/2) & -\sin(\theta/2) \\ \sin(\theta/2) & \cos(\theta/2) \end{bmatrix}$. With this, one can write
\begin{equation}
    |\psi\rangle = R_{z}(\phi)R_{y}(\theta)XP(t/2)X|0\rangle,
\end{equation}
where $P(\eta)=|0\rangle\langle 0|+e^{i\eta}|1\rangle\langle 1|$ is the phase gate.

This building block shall be applied also for more qubits. Let us now regard explicitly the two-qubit case, whose state can be written as follows:
\begin{align}
|\psi\rangle & = \sum_{j,k=0}^{1}c_{jk}|jk\rangle \\
& = \sum_{j=0}^{1}r_{j}|j\rangle r_{j}^{-1}\sum_{k=0}^{1}c_{jk}|k\rangle \\
& = \sum_{j=0}^{1}r_{j}|j\rangle\otimes U_{j}|0\rangle \\
&  = C_{U_0}^{0_0\rightarrow 1}C_{U_1}^{0_1\rightarrow 1}\big(R_y(\xi)|0\rangle\otimes|0\rangle\big)
\end{align}
where $U_j = e^{it_{j}/2}R_{z}(\phi_j)R_{y}(\theta_j)$ with $t_{j}=\phi_{j1}+\phi_{j0}$, $\phi_{j}=\phi_{j1}-\phi_{j0}$, $\phi_{jk} = \arctan(\Im(c_{jk})/\Re(c_{jk}))$, $\theta_j = 2\arctan(|c_{j1}|/|c_{j0}|)$, $\xi=2\arccos(r_0)$, and $r_j^2 = |c_{j0}|^{2} + |c_{j1}|^{2}$. Above we used the controlled unitary $C_{U}^{c_s \rightarrow t}$ with $c$ standing for the control qubit, $t$ is the target qubit and $s$ is the activation state. For instance $C_{U}^{1_0 \rightarrow 0} = U\otimes|0\rangle\langle 0| + I\otimes|1\rangle\langle 1|.$ Besides, we notice that $R_y (\xi)|0\rangle = \sum_{j=0}^{1}r_j |j\rangle$ is a general one-qubit state with real coefficients, when represented in the computational basis (CB). This pattern shall repeat for an $n$-qubit state, that is prepared starting from a $n-1$ state with real coefficients in the CB.

For $3$-qubit states
\begin{align}
|\psi\rangle & = \sum_{j,k,l=0}^{1}c_{jkl}|jkl\rangle  \\
& = \sum_{j,k=0}^{1}r_{jk}|jk\rangle r_{jk}^{-1}\sum_{l=0}^{1}c_{jkl}|l\rangle \\
& = \sum_{j,k=0}^{1}r_{jk}|jk\rangle\otimes U_{jk}|0\rangle \\
& = \Pi_{j,k=0}^{1} C_{U_{jk}}^{0_{j}1_{k}\rightarrow 2}\big(|\Phi\rangle\otimes|0\rangle\big)
\end{align}
with $U_{jk} = e^{it_{jk}/2}R_{z}(\phi_{jk})R_{y}(\theta_{jk})$,  $t_{jk} = \phi_{jk0}+\phi_{jk1}$,
$\phi_{jk} = \phi_{jk1}-\phi_{jk0}$, 
$\theta_{jk} = 2\arctan\big(|c_{jk1}|/|c_{jk0}|\big)$, $\phi_{jkl}=\arctan\big(\Im(c_{jkl})/\Re(c_{jkl})\big)$, $r_{jk}^2 = |c_{jk0}|^2 + |c_{jk1}|^2$ and $|\Phi\rangle = \sum_{j,k=0}^{1}r_{jk}|jk\rangle$ is a two-qubit state with real coefficients.

So, in the general case of $n$ qubits, the state preparation will proceed by the following steps:
\begin{enumerate}
\item Prepare a $n-1$-qubit state
\begin{equation}
|\Phi\rangle = \sum_{j_0,j_1,\cdots,j_{n-2}=0}^{1}r_{j_0,j_1,\cdots,j_{n-2}}|j_0,j_1,\cdots,j_{n-2}\rangle
\end{equation}
with real coefficients 
\begin{equation}
r_{j_0,j_1,\cdots,j_{n-2}} = \sqrt{|c_{j_0,j_1,\cdots,j_{n-2},0}|^{2} + |c_{j_0,j_1,\cdots,j_{n-2},1}|^{2}}.
\end{equation}

\item Apply $2^{n-1}$ multi-controlled unitary gates to $|\Phi\rangle\otimes|0\rangle$ with the first $n-1$ qubits as the control register and the last qubit as the target:
\begin{equation}
\Pi_{j_0,j_1,\cdots,j_{n-2}=0}^{1}C_{U_{j_0,j_1,\cdots,j_{n-2}}}^{0_{j_0},1_{j_1},\cdots,(n-2)_{j_{n-2}}\rightarrow n-1}
\end{equation}
with
\begin{align}
& U_{j_0,j_1,\cdots,j_{n-2}} = \nonumber \\
& e^{it_{j_0,j_1,\cdots,j_{n-2}}/2}R_{z}(\phi_{j_0,j_1,\cdots,j_{n-2}})R_{y}(\theta_{j_0,j_1,\cdots,j_{n-2}}),
\end{align}
where
\begin{align}
& t_{j_0,j_1,\cdots,j_{n-2}} = \phi_{j_0,j_1,\cdots,j_{n-2},1} + \phi_{j_0,j_1,\cdots,j_{n-2},0}, \\
& \phi_{j_0,j_1,\cdots,j_{n-2}} = \phi_{j_0,j_1,\cdots,j_{n-2},1} - \phi_{j_0,j_1,\cdots,j_{n-2},0}, \\
& \theta_{j_0,j_1,\cdots,j_{n-2}} = 2\arctan\left(\frac{|c_{j_0,j_1,\cdots,j_{n-2},1}|}{|c_{j_0,j_1,\cdots,j_{n-2},0}|}\right), \\
& \phi_{j_0,j_1,\cdots,j_{n-1}} = \arctan\left(\frac{\Im(c_{j_0,j_1,\cdots,j_{n-1}})}{\Re(c_{j_0,j_1,\cdots,j_{n-1}})}\right).
\end{align}
\end{enumerate}

It is worth observing that for preparing states with real coefficients in the CB, we use the same algorithm but with
\begin{equation}
U_{j_0,j_1,\cdots,j_{n-2}} = R_{y}(\theta_{j_0,j_1,\cdots,j_{n-2}}).
\end{equation}



\end{document}